\documentclass[]{rspublic}
  
\usepackage[]{graphicx}
\usepackage{lscape}
\usepackage{natbib} 

\usepackage{multirow}
\usepackage{epstopdf}

\begin{document}
\title{Galactic planetary science}
\author[G. Tinetti]{Giovanna Tinetti,$^1$}

\affiliation{ $^1$ Department of Physics and Astronomy, University College London, Gower Street, London WC1E 6BT, UK}
\maketitle
\label{firstpage}
\begin{abstract}{exoplanets; atmospheric models; space missions}
Planetary science beyond the boundaries of our Solar System is today in its infancy. Until a couple of decades ago, the detailed investigation of the planetary properties was restricted to objects orbiting inside the Kuiper Belt. Today we cannot ignore that the number of known planets has increased by two orders of magnitude, nor that these planets resemble anything but the  objects present in our own Solar System. Whether this fact is the result of a selection bias induced by the kind of techniques used to discover new planets -- mainly radial velocity and transit -- or simply the proof that the Solar System is a rarity in the Milky Way, we do not know yet. What is clear, though, is that the Solar System has failed to be \textit{the} paradigm  not only in our Galaxy but even ``just'' in the solar neighbourhood. This finding, although unsettling, forces us to reconsider our knowledge of planets under a different light and perhaps question  a few of the theoretical pillars on which we base our current ``understanding''.

The next decade will be critical to advance in  what we should perhaps call \emph{Galactic planetary science}. In this paper, we review highlights and pitfalls of our current knowledge of this topic and elaborate on how this knowledge might arguably evolve in the next decade. More critically, we identify what should be the mandatory scientific and technical steps to be taken in this fascinating journey of remote exploration of planets in our Galaxy. 
\end{abstract}
\section{Introduction}

If we had to select a single word to define the field of exoplanets, that word would be \emph{revolutionary}.
During the past years, over one thousand planets have been found around every type of stars from A to M, including pulsars and binaries. Being the leftover of the stellar formation processes, planets appear to be rather ubiquitous and, in reality, the presence of a host star is not even a mandatory circumstance. The current statistical estimates indicate that, on average, every star in our Galaxy hosts at least one planetary companion \citep{Cassan2012}, i.e.  our Milky Way is crowded with one hundred billion planets. 

The most revolutionary aspect of this young field is the discovery that the Solar System does not appear to be the paradigm in our Galaxy, but rather one of the many possible configurations we are seeing out there. These include planets completing a revolution in less than one day, as well as planets orbiting two stars or moving on trajectories so eccentric to resemble comets. This variety of stellar and orbital parameters convert into planetary temperatures which span over two orders of magnitudes. 
Unexpectedly, planetary sizes and masses do not appear to be ``quantised" as it happens in the Solar System, where the terrestrial planets are well separated from Neptune and Uranus and those are, in turn, well distinct from Jupiter and Saturn. Instead, a continuum of sizes and masses appear to exist, from the super-Jupiter down to the sub-Earth objects \citep{Batalha2013, Fressin2013}.

While the relative frequency of  ``odd'' planets compared to the ``normal'' ones -- assuming the Solar System planets represent the normality -- might be the result of some selection effects caused by the detection techniques used so far -- mainly radial velocity and transit -- it is undoubtable that 
a great diversity of  planets does exist around other stars. 
In the short term we should be able to shed light on this issue. The European Space Agency's GAIA mission is expected to find several thousands new planets through astrometry \citep{Sozzetti2011, Sozzetti2013}, a technique sensitive to planets lying in a different region of the parameter space compared to transit and radial velocity, in particular to planets at intermediate separation -- typically a few AUs -- from their mother star. 
The instruments ESO-VLT SPHERE \citep{Beuzit2008}, Gemini Planet Imager \citep{Hartung2013} and SUBARU SCExAO \citep{Jovanovic2013} were built to detect young, massive planets at large separation from the stars, a regime not yet well explored till now.

\begin{figure}[h]
   \begin{center}
   \includegraphics[width=0.8\textwidth]{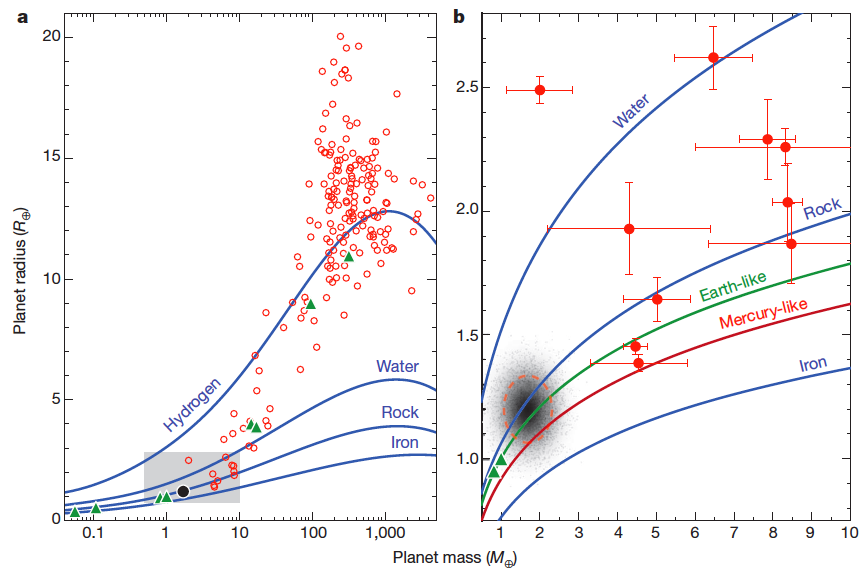} 
 \caption{\small  Masses and radii of currently confirmed transiting planets 
 \citep{Howard2013}. Extrasolar planets
are denoted by red circles and Solar System planets are represented by green triangles. Green and brown lines denote Earth-like composition
(67\% rock, 33\% iron) and Mercury-like composition (40\% rock, 60\% iron).  }
 \label{mr}    
   \end{center}
   \end{figure}

With these numbers and premises, emphasis in the field of exo-planetary science must shift from discovery to understanding: understanding the nature of exo-planetary bodies, and understanding their history.  
The following fundamental questions need to be addressed:
\begin{itemize}
\item	What is the origin of the observed exoplanet diversity?
\item	How and where do exoplanets form?
\item	What are the physical processes  responsible for exoplanet evolution?   
\end{itemize}
In all disciplines, taxonomy is often the first step toward understanding, yet we do not have to date  even a simple taxonomy of planets and planetary systems.
For planets transiting in front of their parent stars -- of which over 400 are known today -- the simplest observables are the planetary radius and, when combined with radial velocity, the mass. Mass and radius allow to estimate the planetary bulk density. From Fig. \ref{mr} it is evident that even gas giants have a broad range of interior structures and core composition, as shown from the different bulk densities observed \citep{Guillot2005, Fortney2007}. While this has stimulated very interesting theoretical work on planetary interior and equations of state of hydrogen at high pressure and temperature, the implications on e.g. planetary formation and evolution mechanisms is still unclear. Most likely, the different bulk densities reflect the different nature and size of the planet's core, which in turn will depend on both the formation mechanism and the ``birth distance'' from the parent star. 
Objects lighter than ten Earth masses (super-Earths/sub-Neptunes, right-hand panel of Fig. \ref{mr}) are even more enigmatic, as they often can be explained in different ways \citep{Adams2008, Grasset2009, Valencia2013}. Among those, Kepler-10 b, Kepler-78 b, CoRoT-7 b and 55 Cnc e all have high densities and orbit G-stars like our Sun with periods of less than one day. By contrast, GJ 1214 b, Kepler-11 d, e, f have much lower densities and are subject to less intense insolation because of their longer period or cooler parent star. In the next years, dedicated space missions such as NASA TESS \citep{Ricker2010} and ESA Cheops \citep{Cheops} combined with radial velocity surveys, will measure the sizes and masses of a few thousands new planets, completing the current statistics of available planetary densities in the solar neighbourhood down to the terrestrial regime. 

\begin{figure}[h]
   \begin{center}
   \includegraphics[width=0.55\textwidth]{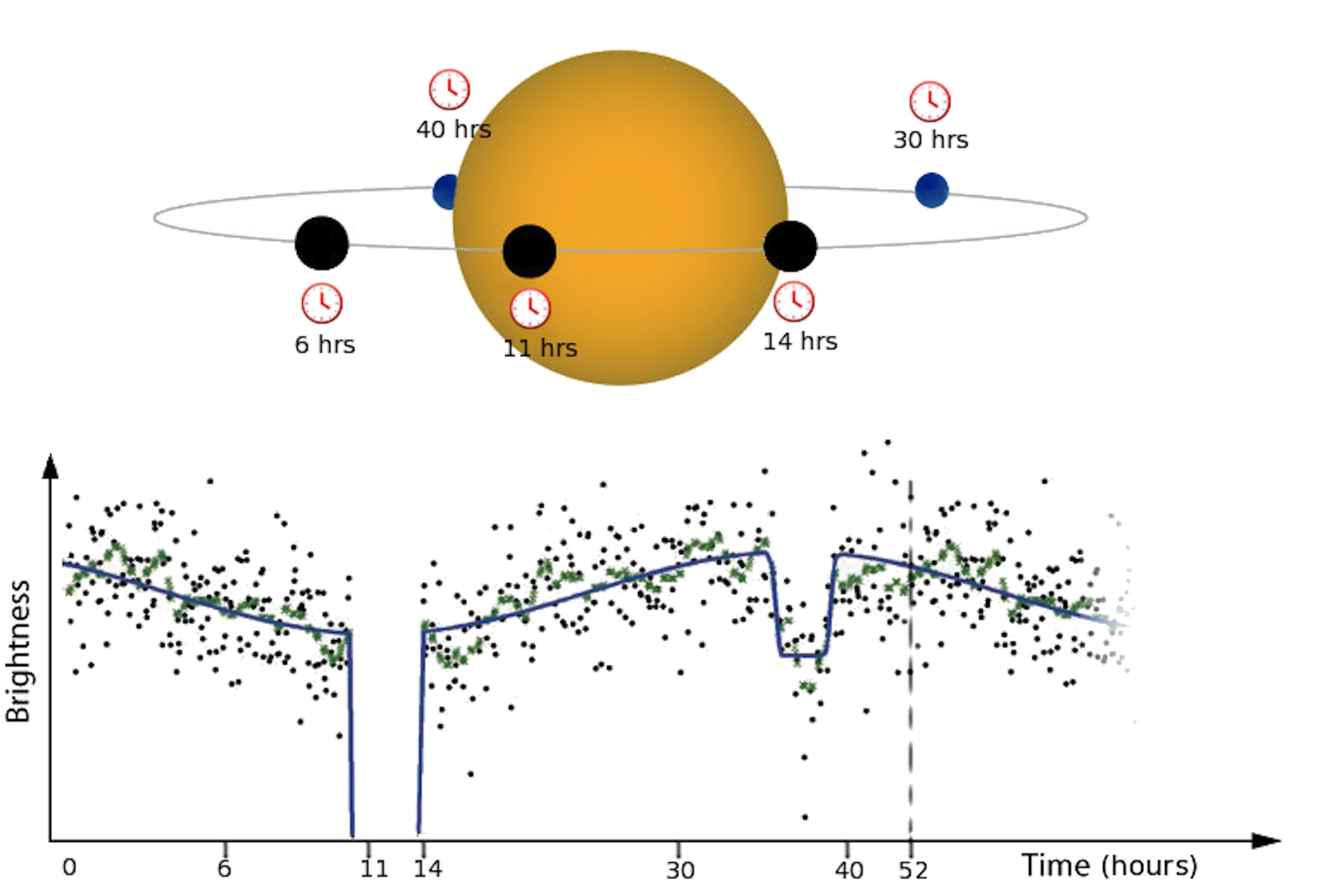} \includegraphics[width=0.44\textwidth,trim=23cm 0cm 0cm 0cm, clip=true]{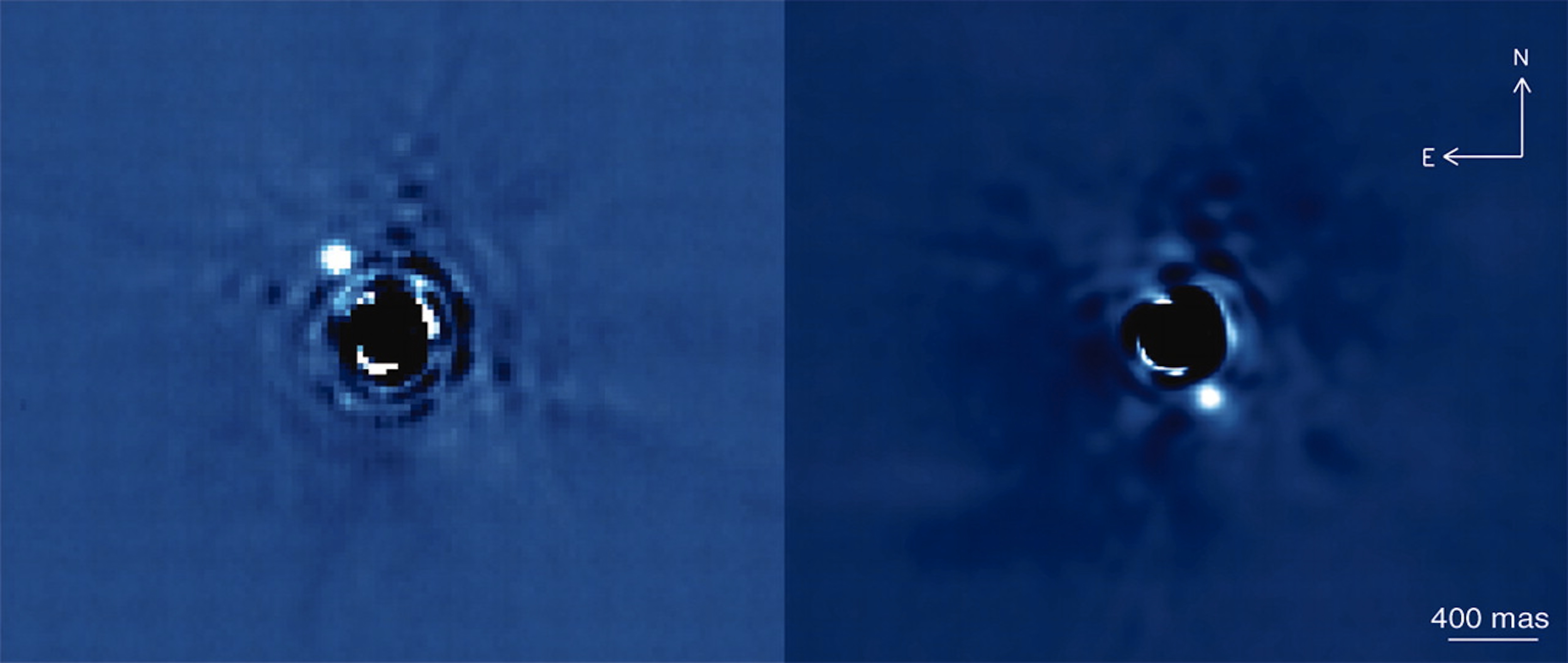}
  \caption{ Left: phase curve of the hot-Jupiter HAT-P7b while orbiting around its mother star as observed by Kepler \citep{Borucki2009}. The transit and eclipse events 
 are occurring at   10h and 35h respectively. Right: image of the planet $\beta$-Pic b, located 8 to 15 astronomical units from the star, as observed with instrument VLT-Naco \citep{Lagrange2010}. 
   }  \label{transit} 
   \end{center}
   \end{figure}

As explained above, density is a very important parameter, but alone it cannot be used as a discriminant of the variety of cases we are seeing out there. We need additional information to proceed. 
The other key observable for planets is the chemical composition and state of their atmosphere. Knowing what are atmospheres made of is essential to clarify, for instance, whether a planet was born in the orbit it is observed in or whether it has migrated a long way;  it is also critical to understand the role of stellar radiation on escape processes, chemical evolution and global circulation. To date two methods can be used to sound exoplanetary atmospheres: transit and eclipse spectroscopy and direct imaging spectroscopy. These are very complementary methods 
 and we should pursue both to get a coherent picture of planets outside our solar system (Fig. \ref{transit}). 
 
\section{Brief review of exoplanet spectroscopic observations}
\subsection{Transit}
When a planet passes in front of its host star (transit), the star flux is reduced by a few percent, corresponding to the planet/star projected area ratio (transit depth). The planetary radius can be inferred from this measurement. If atomic or molecular species are present in the exoplanet's atmosphere, the inferred radius is larger at some specific wavelengths (absorption) corresponding to the spectral signatures of these species \citep{Seager2000, Brown2001, Tinetti2007b}. The transit depth $\kappa (\lambda)$  as a function of wavelength ($\lambda$) is given by: 
													\begin{equation}
\kappa(\lambda) = \frac{R_p^2 + 2\, \int_{0}^{z_{max}} (R_p+z)\, (1-\text{e}^{- \tau(z,\lambda)}) \text{d}z}{R_*^2}
\label{eq1}
\end{equation}
where $R_*$ is the stellar radius, $z$ is the altitude above $R_p$ and $\tau$ the optical depth. Equation (\ref{eq1}) has a unique solution provided we know $R_p$ accurately. $R_p$ is the planetary radius at which the planet becomes opaque at all $\lambda$. For a terrestrial planet, $R_p$ usually coincides with the radius at the surface. In contrast, for a gaseous planet, $R_p$ may correspond to a pressure $p_0 \sim$ 1-10 bar, depending on the transparency of the atmosphere.
\subsection{Eclipse} A  measurement of the planet's emission/reflection can be obtained through the observation of the planetary eclipse, by recording the difference between the combined star + planet signal, measured just before and after the eclipse, and the stellar flux alone, measured during the eclipse. In contrast with the primary transit observations, the dayside of the planet is observed, which makes both methods fully complementary. Observations provide measurements of the flux emitted and/or reflected by the planet in units of the stellar flux \citep{Charbonneau2005, Deming2005}. The planet/star flux ratio $\phi (\lambda)$ is defined as: 
\begin{equation}
                                                         \phi (\lambda) = \frac{R_p^2}{R_*^2}  F_p (\lambda)/F_* (\lambda)		
                                                         \label{eq2}		          
\end{equation}
\subsection{Phase curves}
In addition to transit and eclipse observations, monitoring the flux of the star+planet system over the orbital period allows the retrieval of information on the planet emission at different phase angles. Such observations have to be performed from space, as they typically span over a time interval of more than a day \citep{Harrington2006, Knutson2007, Borucki2009, Snellen2010}.
\subsection{Direct imaging} 
The planet to star brightness contrast may typically range  between 10$^{-4}$ and 10$^{-10}$ depending on many parameters of the system, i.e. age, distance, planetary size, temperature etc. and of course spectral interval.  To fix the ideas, Jupiter has a contrast of about 10$^{-9}$ relative to the Sun in the visible  and an angular separation of 0.5'' at 10pc. 
The use of a coronagraphic system \citep{Lyot1939, Guyon2013} is therefore essential to extract the planetary signal out of the stellar light. 
\\
Wavefront aberrations and stellar speckles are another critical problem that needs to be attenuated. Deformable mirrors \citep{Trauger2007} and speckle calibration techniques,  such as angular differential imaging  \citep{Marois2008}, can be used effectively to address this issue.
\section{Highlights \& problems with current  photometric and spectroscopic data}
\subsection{Highlights}
Water vapour appears to be ubiquitous in the atmospheres of transiting hot-Jupiters with temperatures between 800 and 2200 K observed to date \citep{Barman2007, Tinetti2007, Grillmair2008, Beaulieu2010, Swain2008, Crouzet2012, Deming2013, Birky2013}. The additional presence of carbon-bearing species, such as methane, carbon monoxide and dioxide in those atmospheres has been supported by both observations and spectral simulations \citep{Swain2008, Swain2009, Swain2009b, Tinetti2010,  Snellen2010, Brogi2012, deKok2013}, but their relative abundances are still unclear \citep{Swain2009b, Madhusudhan2009, Lee2012, Line2012, Tinetti_faraday}. Nitrogen-bearing species -- e.g. HCN, NH$_3$ -- are most probably also there \citep{Moses2011, Venot2012}, but current observations are not precise enough to indicate their presence.
Ground-base observations in the L-band have  been interpreted as bearing the signature of methane fluorescence in the atmosphere of one of these hot-jupiters   \citep{Swain2010, Waldmann2012}, 
this would be an important diagnostic of the physical structure of the upper atmosphere of these planets probed through a minor atmospheric constituent. In the atmosphere of very hot-Jupiters, where temperatures approach 3000 K, exotic species commonly present in brown dwarfs, such as  metal oxides (TiO, VO) or metal hydrides (CrH, TiH etc.) have been suggested to explain observations by the Hubble-STIS and WFC3 \citep{Barman2007, Swain2013, Stevenson2013}. These species are important as they may influence both the planetary albedo and the vertical thermal structure of the planet. 
Sodium and perhaps potassium, are present in most hot-Jupiters analysed \citep{Charbonneau2002, Redfield2007, Snellen2008, Colon2011}. Apart from these alkali metals, the spectra in the visible  appear dominated by Rayleigh scattering or condensates/hazes \citep{Knutson2007a, Sing2011}. 
\\
Warm Neptunes are expected to be methane-rich \citep{Line2010, Moses2011, Venot2013}, and indeed photometric observations of GJ 436b may point in this direction \citep{Beaulieu2011}. Spectroscopy will be needed to unravel the full picture of this and other objects, such as GJ 3740b \citep{Fukui2013, Nascimbeni2013}. The $\sim$6 Earth-mass, warm planet GJ 1214b is the first super-Earth which has been probed spectroscopically \citep{Bean2010}. The VLT observations were followed by other space and ground observations \citep{Berta2012} that are suggestive of an atmosphere heavier than pure molecular hydrogen, but additional observations are needed to confirm its composition \citep{Kreidberg2014}. \\
Information on the stability of the atmospheres of transiting planets has been collected through UV observations with Hubble \citep{Vidal-Madjar2003, Linsky2010, Fossati2010}: hydrodynamic escape processes are likely to occur for most of the planets orbiting too close to their parent star \citep{Yelle2004, Koskinen2007, Garcia2007, Koskinen2013a, Koskinen2013b}. 
Also, orbital phase curves in the IR \citep{Harrington2006, Knutson2007, Crossfield2010} and eclipse mapping measurements \citep{Majeau2012, deWit2012} have provided first constraints on the thermal properties and dynamics of hot-Jupiters' atmospheres.

In parallel with transit studies, in the next decade direct imaging techniques are expected to allow observations of hot, young planets at large separations from their parent star, i.e. gaseous planets newly formed in the outer regions of their planetary disc and not (yet?) migrated inward. \\
Multiple-band photometry and spectroscopy
in the near-infrared (1-5 $\mu$m) have been obtained for a few young gaseous planets, such as $\beta$ Pic-b  \citep{bon13,cur13}, GJ 504 b \citep{jan13} and the planets around HR 8799 \citep{kon13}.
These observations will be perfected and extended to tens of objects with dedicated instruments such as SPHERE and GPI. The comparison of the chemical composition of these young gaseous objects with the composition of their migrated siblings probed through transit, will enable us to understand the role played by migration and by extreme irradiation on gaseous planets.

\subsection{Issues \& possible solutions}
Although the field of exoplanet spectroscopy has been very successful in the past years, there are a few serious hurdles that need to be overcome to progress in this area, in particular:
\begin{enumerate}
\item \emph{  Instrument systematics are often difficult to disentangle from the signal. } In the past, parametric models have extensively been used by most teams to remove instrument systematics. This approach has caused many debates regarding the use of different parametric choices to remove the systematic errors. Parametric models approximate systematic noise by fitting a linear combination of optical state vectors to the data (e.g. X and Y-positional drifts of the star or the spectrum on the detector, the focus and the detector temperature changes, positional angles of the telescope on the sky). Even when the parameterisation is sufficient, it is often difficult to determine which combination of these parameters may best capture the systematic effects of the instrument. \\
Unsupervised machine learning algorithms do not need to be trained prior to use and do not require auxiliary or prior information on the star, instrument or planet. The machine learning approach will ``learn'' the characteristics of an instrument from observations, allowing to de-trend systematics from the astrophysical signal. This approach guarantees a higher degree of objectivity compared to traditional methods. In \citet{Waldmann2012, Waldmann2013, Waldmann2013b, Morello2013} Independent Component Analysis (ICA; \cite{Hyvarinen1999}) has been adopted as an effective way to decorrelate the exoplanetary signal from the instrument systematics in the case of Hubble-NICMOS and Spitzer/IRS-IRAC data.
\item \emph{  Especially for transit observations, stellar activity is the largest source of astrophysical noise. }
Stellar noise is an important source of spectral and temporal instability in exoplanetary time series measurements \citep{Agol2008, Ballerini2012}. This is particularly true for M dwarf host stars as well as many non-main sequence stars. Correction mechanisms for fluctuations must be an integral part of the data analysis. The problem of stellar activity removal from time series data is a very active field of research. Whereas most instrumental effects can be measured or calibrated to some degree, stellar and general astrophysical noise does not usually grant us this luxury. 
In \cite{Waldmann2012, Danielski2013} the same methods explained in point 1.  to decorrelate the systematic noise, were successfully used to eliminate/reduce the effects of the stellar activity in Kepler photometric light-curves. These methods need  to be applied  to spectroscopic time-series, to assess their validity and potential also in the spectral domain.
\item \emph{ Data are sparse, i.e. there is not enough wavelength coverage and most of the time the observations were not recorded simultaneously. }
\item \emph{ Absolute calibration at the level of $10^{-4}$ is not guaranteed by current instruments, therefore caution is needed when one combines multiple datasets not recorded simultaneously. }
\item \emph{ Transmission and emission spectra, as measured through transit, eclipse and direct imaging, are intrinsically degenerate}. 
In transit spectroscopy the degeneracy in the retrieval of molecular abundances, may be caused by the imprecise knowledge of $R_p$ (Eq. \ref{eq1}).  
In IR eclipse  and direct imaging spectroscopy, the information on molecular abundances is entangled with the atmospheric vertical thermal profile,
see for instance \citet{Tinetti13} for a more detailed discussion.
For transiting planets, to remove the degeneracy between molecular abundances/planetary radius or molecular abundances/vertical thermal gradient, a broad wavelength coverage is needed 
together with  adequate SNR and spectral resolving power (see point 7). Direct imaging observations also suffer from the lack of knowledge of the planetary radius and sometimes mass.
When the mass and the radius are not known, model estimates need to be invoked, increasing the source of degeneracy.  
\item \emph{ Accurate linelists are an essential element of radiative transfer models, and this fact is not always appreciated. } As a results, the abundances for molecular species  are often derived with linelists which are incomplete or extrapolated from measurements/calculations at low temperature. This issue --especially together with point 3.-- may introduce large errors. For instance all the current claims of carbon-rich or carbon-poor planets \citep{Madhusudhan2012} published in the literature are  unsubstantiated for this reason. 
 This problem is well known to spectroscopists, and linelists at high temperatures are  being calculated ab initio or measured in laboratory \citep{Tennyson2012}.
\item \emph{ We are dealing with very low SNR observations}. While the adoption of new data analysis methods and models might address some of the issues listed above, the lack of good data is something we cannot solve in the short term.
We indicate  below the SNR per spectral resolution element and spectral resolving power (SRP) that would be needed to guarantee a sound spectral retrieval. 
We refer to \cite{Tinetti13, tessenyi13} for a more extensive discussion of these parameters.
\begin{enumerate}
\item \emph{Basic}. SNR$\sim$5 and a SRP $\sim 50$ for $\lambda < 5 \mu$m and $\sim 30$ for $\lambda > 5 \mu$m. Key molecular species can be detected, main thermal properties are captured.
\item \emph{Deep}. SNR$\sim$10 and a SRP $\sim 100$ for $\lambda < 5 \mu$m and $\sim 30$ for $\lambda > 5 \mu$m. Molecular abundances (i.e. the abundance of one component relative to that of all other components) are retrieved and so is the vertical thermal structure.
\item \emph{Ultra-deep}. SNR$\sim$20 and a SRP $\sim 300$ for $\lambda < 5 \mu$m and $\sim 30$ for $\lambda > 5 \mu$m. A very thorough spectral retrieval study can be performed.
\end{enumerate}
In Tables \ref{tab1} and \ref{tab2} we show the detectable molecular abundances at fixed SNR and SRP for a typical warm Neptune, such as GJ 436 b, and a hot super-Earth, such as 55 Cnc e. The results for hot-Jupiters are very similar to the ones reported for warm Neptunes. We refer to \citep{tessenyi13} for the case of temperate super-Earths around late dwarfs. 
\begin{table}
\footnotesize
\begin{tabular}{lcclcclccclccc}
\hline
 			& \multicolumn{2}{c}{$CH_4$} 			&& \multicolumn{2}{c}{$CO$} 		&& \multicolumn{3}{c}{$CO_2$} 						&& \multicolumn{3}{c}{$NH_3$} 				 \\
Retrieval 	& 3.3$\mu m$ & 8$\mu m$ 			&\,& 2.3$\mu m$ & 4.6$\mu m$ 	&\,& 2.8$\mu m$ & 4.3$\mu m$ & 15$\mu m$ 				&\,& 3$\mu m$ & 6.1$\mu m$ & 10.5$\mu m$ 		 \\
\hline
Ultra-deep 	& $10^{-7}$ & $10^{-6}$ 				&&	$10^{-4}$	& $10^{-6}$ 		& & $10^{-7}$ & $10^{-7}$ & $10^{-7}$ 					&& $10^{-7}$ & $10^{-6}$ & $10^{-7}$ 			 \\
Deep	& $10^{-7}$ & $10^{-6}$ 				&&	$10^{-3}$	& $10^{-5}$ 		& & $10^{-6}$ & $10^{-7}$ & $10^{-6}$ 					&& $10^{-6}$ & $10^{-6}$ &$10^{-6}$ 			 \\
basic 	& $10^{-7}$ & $10^{-5}$ 				&&	$10^{-3}$	& $10^{-4}$ 		& & $10^{-6}$ & $10^{-7}$ & $10^{-5}$ 					&& $10^{-5}$ & $10^{-5}$ & $10^{-5}$			 \\
\\
\end{tabular}\\
\begin{tabular}{lcclcclccclccc}
\hline
				& \multicolumn{2}{c}{$PH_3$} 	& & \multicolumn{2}{c}{$C_{2}H_6$}		&& \multicolumn{3}{c}{$H_{2}S$} 				&& \multicolumn{3}{c}{$C_{2}H_2$} \\
Retrieval		& 4.3$\mu m$ & 10$\mu m$ 		&\,& 3.3$\mu m$ & 12.2$\mu m$ 		&\,& 2.6$\mu m$ & 4.25$\mu m$ & 8$\mu m$ 		&\,& 3$\mu m$ & 7.5$\mu m$ & 13.7$\mu m$ \\
\hline
Ultra-deep 		& $10^{-7}$ & $10^{-6}$			&& $10^{-6}$ & $10^{-6}$				&&	$10^{-5}$	& $10^{-4}$ & $10^{-4}$			&& $10^{-7}$ & $10^{-5}$ & $10^{-7}$\\
Deep		& $10^{-7}$ & $10^{-6}$			&& $10^{-5}$ & $10^{-5}$				&&	$10^{-5}$	& $10^{-4}$ & $10^{-3}$			&& $10^{-7}$ & $10^{-4}$ & $10^{-6}$\\
Basic 		& $10^{-7}$ & $10^{-5}$ 		&& $10^{-5}$ & $10^{-5}$				&&	$10^{-4}$	& $10^{-3}$ & -					&& $10^{-7}$ & $10^{-3}$ & $10^{-5}$\\
\hline
\\
\end{tabular}\\
\begin{tabular}{lccclcccl}
\hline
	& \multicolumn{3}{c}{$HCN$} 			&& \multicolumn{3}{c}{$H_{2}O$} \\
Retrieval 	&    3$\mu m$ & 7$\mu m$ & 14$\mu m$    &\,& 2.8$\mu m$ & 5-8$\mu m$ & 11-16$\mu m$  \\
\hline
Ultra-deep 	&   $10^{-7}$ & $10^{-5}$ & $10^{-7}$	&& $10^{-6}$ & $10^{-6}$	&  $10^{-5}$ \\
Deep	& $10^{-6}$ & $10^{-5}$ & $10^{-6}$	&& $10^{-6}$ & $10^{-5}$	& $10^{-4}$ \\
Basic 	&		$10^{-6}$ & $10^{-4}$ & $10^{-5}$ 	&& $10^{-5}$ & $10^{-5}$	& $10^{-4}$ \\
\hline
\\
\end{tabular}\\
\caption{\small Warm Neptune: minimum detectable abundances (mixing ratios) for a basic, deep and ultra-deep retrieval.  The bulk composition of the planetary atmosphere in this simulation is molecular hydrogen with a small fraction of Helium. See \cite{tessenyi13} for further details on the method.}
\label{tab1}
\end{table}
\begin{table}
\small
\begin{tabular}{lccclccc}
\hline
\hline
 			& \multicolumn{3}{c}{$H_{2}O$} 							&  & \multicolumn{3}{c}{$CO_2$} \\	
Retrieval	& 2.8$\mu m$ & 5 - 8$\mu m$ & 11 - 16$\mu m$ 				& \, & 2.8$\mu m$ & 4.3$\mu m$ & 15$\mu m$ \\
\hline
Ultra-deep 	& $10^{-4}$ & $10^{-4}$ & $10^{-4}$						& &	$10^{-5}$	& $10^{-7}$ & $10^{-5}$ \\
Deep		& $10^{-4}$ & $10^{-3}$ &  $10^{-3}$						& &	$10^{-5}$	& $10^{-6}$ & $10^{-4}$  \\
basic 	& $10^{-3}$ 	& - 		&   - 								& &	$10^{-4}$	& $10^{-5}$ & -			 \\
\hline
\\
\end{tabular}
\caption{\small Hot super-Earth, around a G type star: minimum detectable abundances (mixing ratios) for a basic, deep and ultra-deep retrieval. The bulk composition of the planetary atmosphere in this simulation is $H_{2}O$. See \cite{tessenyi13} for further details on the method and the impact of other main atmospheric component on the results.}
\label{tab2}
\end{table}

\end{enumerate}

\section{The next decade and beyond}
In section 3 (b) we identify the hurdles that cannot be solved in the short term (in particular 3., 4., 7.):  a new generation of ground and space facilities is needed to tackle those. 
In next decade, new large, general purpose observatories from space and the ground will come on line, notably JWST and E-ELT:
it is understood that, among many other science goals, they will significantly contribute to exoplanet spectroscopic observations, both in transit and direct imaging \citep{Clampin2010, Snellen2013, Kasper2013}. 
More crucially for this field, dedicated instruments and missions are  being studied or planned.

The idea of a dedicated IR observatory in space to study exoplanetary atmospheres is clearly not new: back in the eighties
Bracewell (Nature, 1978) and Angel et al. (Nature,
1986) proposed that exoplanets around nearby stars
could be detected in the IR (6-17 $\mu$m) and their spectra
analysed, searching for CO$_2$, H$_2$O, O$_3$, CH$_4$, and
NH$_3$ spectral features. The proposal to implement this idea
under the form of an IR nulling interferometer in space
came almost ten years later \citep{leger96}. The
concept, named DARWIN, was first proposed to ESA in
1993, when the only known planets were the nine in our
Solar System (+ three around a neutron star). Its principal
objectives were to  detect Earth-like planets
around nearby stars, to analyse the composition of their
atmospheres and assess their ability to sustain life as
we know it. Similar mission concepts were proposed to NASA in the US (Terrestrial Planet Finder-Interferometer, \cite{tpfi}). The working hypothesis
of an Earth-twin + Sun-twin as the only cradle for
life, was too geocentric to survive the ``exoplanet revolution'' and none of these very challenging missions have been implemented.
A couple of decades of exoplanet discoveries
have taught us that the pathways to habitable
planets are multiple, but the most interesting ones are those able to cast light on a host of physical and
chemical processes not entirely understood or missing altogether
in our Solar System (Blue Dots report \footnote{http://www.blue-dots.net/sites/blue-dots/IMG/pdf/BD\_report\_V2-02.pdf} \citep{Pathways2010}).

\begin{figure}[h]
   \begin{center}
      \center
   \includegraphics[width=0.9\textwidth]{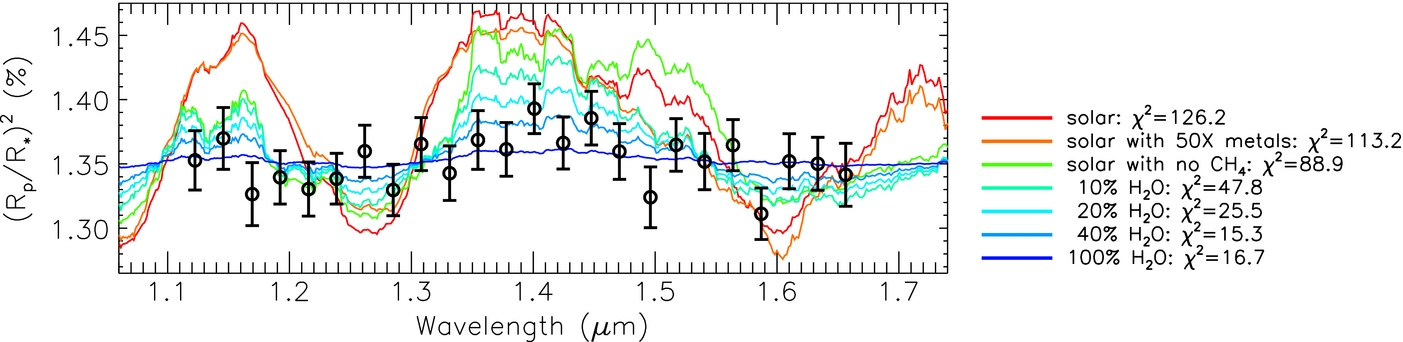} \includegraphics[width=0.7\textwidth]{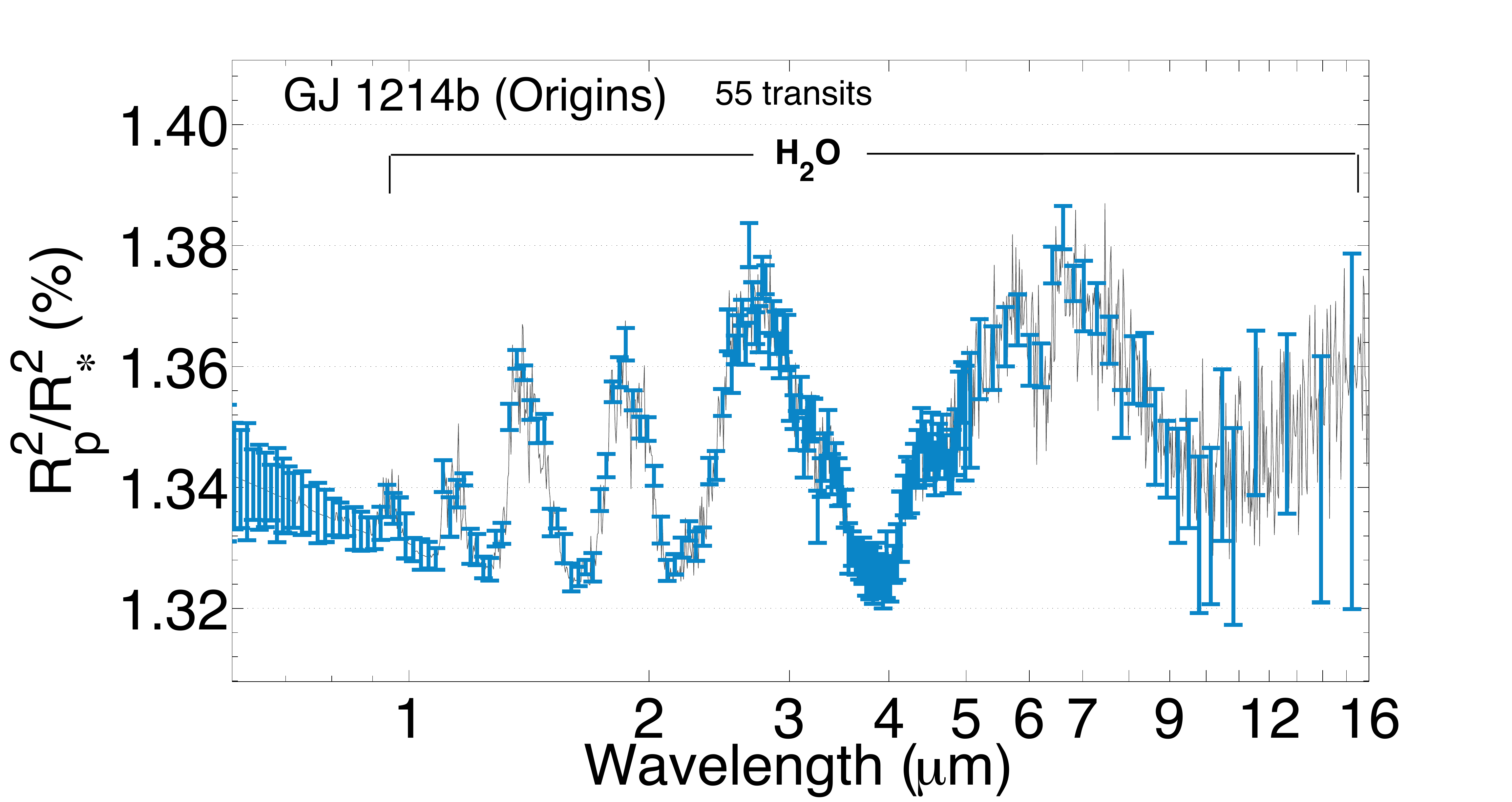}
   \caption{ Top: transit spectroscopic observations  of the super-Earth GJ1214 b recorded with  Hubble-WFC3 \citep{Berta2012}. Bottom:
   simulations of EChO performances compatible with a ``Deep retrieval''.  In December 2013, about 160 known exoplanets could be observed with SNR and SRP corresponding to ``Basic", a good fraction of them as ``Deep", and about 10 to 20 as ``Ultradeep", see (http://sci.esa.int/echo/) and 
   \citet{Varley2014}.    } \label{fig3}
   \end{center}
   \end{figure}

In the past years mission concepts for IR transit spectroscopy from space were proposed to and studied by both ESA and NASA, in particular THESIS \citep{Swain_thesis_2009}, Finesse \citep{Deroo2012} and EChO
(\citep{tinetti2012}).
The transit and eclipse spectroscopy methods allow us to measure atmospheric signals from the planet at levels of at least $10^{-4}$ relative to the star. 
No angular resolution is needed, as the signals from the star and from the planet are differentiated using the knowledge of the planetary ephemerides. 
This can only be achieved in conjunction with a carefully designed, stable payload and satellite platform. 
EChO, the Exoplanet Characterisation Observatory, is currently one
of the five M3 mission candidates  being assessed by ESA, for
a possible launch in 2022 \footnote{http://sci.esa.int/science-e/www/area/index.cfm?fareaid=124}.  
If selected, EChO will provide low-mid resolution (30 to 300), simultaneous multi-wavelength spectroscopic observations (0.55-11 $\mu$m, goal 0.4-16 $\mu$m)  of a few hundred planets, including hot, warm and temperate
gaseous planets and super-Earths around different stellar types. These measurements will allow the
retrieval of the molecular composition and thermal structure of the atmospheres observed. 
The design of the whole detection chain and satellite will be optimised to achieve a high degree of photometric stability (i.e. $\sim$ 100 ppm in 10 hours) and repeatability:
the telescope will have a collecting area of 1.13 m$^2$,  will be diffraction limited at 3 $\mu$m and will be positioned at L2. This Lagrangian point  provides a cold and stable thermal environment, as well as a large field of regard to allow efficient time-critical observation of targets randomly distributed over the sky.
We show in Fig. \ref{fig3}  the simulated performances achievable by EChO to observe the warm super-Earth GJ 1214 b.  Planets which are much smaller (less than 1.5 Earth radii) and colder than this one (colder than 300 K), will  be challenging for an EChO-like mission. Temperate super-Earths may be observable only around bright late M dwarfs. 

Provided a small/medium size transit spectroscopy mission is launched in the next decade, would it make sense to envisage a large spectroscopy mission later on? Probably not.  To illustrate why, it is useful to first discuss  a few basic concepts. 
The   number of electrons per spectral element on the detector   from the star ($N_*$) and planet ($N_p$)  are:
\begin{equation}
N_* = \frac{\Delta F_* \cdot A_{\text{eff}} \cdot \eta \cdot Q}{n} \cdot t; \qquad \qquad 
N_p = \phi \cdot N_* = \phi \cdot \frac{\Delta F_* \cdot A_{\text{eff}} \cdot \eta \cdot Q}{n} \cdot t
\end{equation}
$\phi$ is the planet to star contrast defined in (\ref{eq2}), $\Delta F_*$ is the stellar flux in the spectral band observed (photons/s/m$^2$), $A_{\text{eff}}$ is the effective collecting area (m$^2$),
$\eta$ is the instrumental throughput (dimensionless), $Q$ is the detector quantum efficiency (e$^-$/photons) and $t$ is the integration time (s).
If we assume the observations to be dominated by the stellar photon noise, the SNR per spectral element is:
\begin{equation}
SNR = \frac{N_p}{\sqrt{N_*}} = \phi \cdot \sqrt{\frac{\Delta F_* \cdot A_{\text{eff}} \cdot \eta \cdot Q \cdot t }{n}}
\end{equation}
The SRN thus scales with $\sqrt{A_{\text{eff}}}$, i.e. it goes linearly with telescope diameter ($D$). The cost of a telescope scales, in the most optimistic cost models, as $D$ to the 1.2 power, with some models indicating an exponent of 2 \citep{Stahl2012}. Therefore an increase in telescope diameter of a factor of 2 means a cost increase of a factor 2 to 4, while doubling the SNR has a small to negligible impact on the science.
\\
To be transformational, we should aim at an improvement of at least a factor 10 in the SNR, and this would require to abandon the idea of an agile, highly stable platform  
in favour of  a large, deployable structure, as monolithic space telescopes are limited by fairing size to about 4 m diameter.
 Said structure might represent an encumbrance when trying to reach the pointing stability required by transit observations and certainly might limit the ability to move and repoint agilely from one target to another in the sky.
Note that a factor 10 in SNR might not be sufficient in any case to observe the atmospheres of Earth-like planets around Sun-like stars. For those targets, in fact, transits are expected to occur once per year, 
and 5-6 transits (assuming a mission lifetime of $\sim$5 years) will not be enough to collect the required photons.

Direct imaging from space is the expected next step to be taken in space after transit. Space telescopes with various types of coronagraphs are being studied in the US, Europe and Japan \citep{Clampin2010a, Cash2010, Kasdin2012, Guyon2013a, Serabyn2013, Boccaletti2013, Enya2011}.
A mission for direct imaging, would be technically more challenging than a transit one and certainly more expensive -- the telescope cannot be a light bucket, to start with.
Said mission, though, would open the spectroscopic exploration of planets at larger separation from the stars, a domain which is impracticable with transits. 

In the past two decades the field of exoplanets has spoiled us in terms of creativity and transformational ideas, so perhaps we should not be too surprised if a new technology or a new observing strategy  comes online soon, 
making all the other techniques obsolete or just inefficient.

\section{Acknowledgements}
GT is a Royal Society University Research Fellow.
\bibliographystyle{rspublicnat}
\bibliography{jt,tinetti}

\end{document}